\documentclass[reprint,pra]{revtex4-1}
\usepackage{geometry}                
\usepackage{graphicx}
\usepackage{amssymb}
\usepackage{amsmath}
\usepackage{epstopdf}
\usepackage{color}

\begin{document}
\title{Fisher information vs. signal-to-noise ratio for a split detector}
\author{George~C.~Knee}
\email{gk@physics.org}
\author{William~J.~Munro}
\affiliation{NTT Basic Research Laboratories, NTT Corporation, 3-1 Morinosato-Wakamiya, Atsugi, Kanagawa 243-0198, Japan}
\date{\today}                                           
\begin{abstract}
\noindent
We study the problem of estimating the magnitude of a Gaussian beam displacement using a two-pixel or `split' detector. We calculate the maximum likelihood estimator, and compute its asymptotic mean-squared-error via the Fisher information. Although the signal-to-noise ratio is known to be simply related to the Fisher information under idealised detection, we find the two measures of precision differ markedly for a split detector. We show that a greater signal-to-noise ratio `before' the detector leads to a greater information penalty, unless adaptive realignment is used. We find that with an initially balanced split detector, tuning the normalised difference in counts to $0.884753\ldots$ gives the highest posterior Fisher information, and that this provides an improvement by at least a factor of about 2.5 over operating in the usual linear regime. We discuss the implications for weak-value amplification, a popular probabilistic signal amplification technique.
\end{abstract}
\maketitle

\section{Introduction}
Many historic experiments have concerned the detection of a spatial or angular deflection of a beam of particles or light: examples include Stern and Gerlach's discovery of spin angular momentum~\cite{GerlachStern1922}, Young's double slit experiment~\cite{Young1804} and Germer and Davisson's demonstration of electron diffraction~\cite{Davisson1928}. This tradition continues in investigations of the spin-Hall effect of light~\cite{HostenKwiat2008} and other optical phenomena~\cite{AielloWoerdman2008}. Often the effect being observed is very subtle, and the use of precision instrumentation is necessary to estimate its magnitude.  Charged coupled devices (CCDs) or CMOS 	Active Pixel Sensor arrays are affordable solid-state technologies found in commercial cameras, which typically feature pixel counts in the tens of millions. On the other hand, the use of various detection systems in the field of quantum imaging, including a single pixel camera~\cite{SunEdgarBowman2013}, has showcased the flexibility of few-pixel based detection methods when teamed with clever illumination strategies. In this work we consider how information is lost in one such imperfect detection process: the amount of information `coming out' of a split detector versus the amount of information `going in'.

A split detector is a popular detection system in optics, microscopy, and imaging: it is composed only of a pair of photodiodes, arranged so that a lateral displacement of a beam of particles impinging upon the detector induces a modulation in the relative intensity of the photocurrents from each diode. Although here it is couched in optical terminology, the idea of split detection or `binned' data applies to a wide range of experimental scenarios. 

In this article, we study the ability of such a device to enable estimation of the magnitude of a beam displacement, when lateral beam profile is described by a Gaussian function. In terms of data processing, we consider a simple linear estimator as well as the maximum likelihood estimator, and calculate two measures of precision: Fisher information and signal-to-noise. Our aim is to compare these quantities and their meaning. 

In Sec.~\ref{beam_displacement} we introduce our model of beam displacement, and calculate both the signal-to-noise ratio and Fisher information given ideal detection. In Sec.~\ref{high_res} we introduce the idea of pixelated or `binned' data, and take the limit of poor resolution in Sec.~\ref{split} where our analysis of split detection begins in earnest. There we calculate the posterior Fisher information. This quantity describes the performance of the maximum likelihood estimator, which we derive in  Sec.~\ref{crb}. In Sec.~\ref{split_snr} we derive the signal-to-noise ratio of a split detector and show that it is not (in general) a good measure of precision. Sec.~\ref{wva} applies our results to weak-value amplification, which is a probabilistic method for increasing signal-to-noise. We draw several conclusions in Sec.~\ref{discussion}, including a short discussion on optimal use of split detectors.

\section{Beam displacement}
\label{beam_displacement}
Consider a beam of light, or particles, with a lateral intensity profile described by a Gaussian function centred on $x_0$ with standard deviation $\sigma$:
\begin{align}
p(x) = \frac{1}{\sqrt{2\pi}\sigma}\textrm{e}^{-(x-x_0)^2/2\sigma^2}.
\label{gaussian_model}
\end{align}
The intensity could represent flux of quanta of energy, or simply the amplitude of the electromagnetic field -- here we model it as a probability density function. A displacement of the beam is then simply written as $p'(x)=p(x-d)$; it might be generated (for example) by a magnetic field, an interference effect, or through coupling to an interface in the propagation medium. It will be convenient to consider  $d = \lambda g$, where $\lambda$ is a fixed and known parameter and $g$ is an unknown parameter of interest, and the subject of a parameter estimation study.

If one is ignorant of the value of $g$, but permitted a large number of samples from $p(x-\lambda g)$,  one may venture a guess, or estimate, of the true value. Knowledge of the functional form of $p'(x)$ allows for different philosophies to inform the estimate. 
Maximum likelihood estimation (MLE) is one of the most powerful estimation strategies. Under reasonable conditions~\cite{Wolfowitz1965,Wald1949}, it provides an estimate $\tilde{g}_{\text{\tiny MLE}}$ for the unknown parameter that i) is \emph{unbiased} $\mathbb{E}( \tilde{g}_{\text{\tiny MLE}} ) = g$ ($\mathbb{E}$ denotes the expectation value), and ii) is \emph{efficient} , that is to say has a mean squared error that decays (when the number of trials $N\rightarrow\infty$) as $\mathbb{E}((\tilde{g}_{\text{\tiny MLE}}-g)^2) = \text{Var}(\tilde{g}_{\text{\tiny MLE}})=1/(FN)$. The proportionality constant $F$ is the Fisher information, which depends on $p'(x|g)$. The asymptotic mean squared error cannot be beaten by any other unbiased estimation strategy, and thus MLE is said to saturate the Cram\'er-Rao bound~\cite{Van-Trees1968}. 

The likelihood of $g$ conditioned on $N$ independent samples $x_i$ is 
\begin{align}
\mathcal{L}(g|x_i)=\prod_i^N p'(x_i|g),
\end{align}
and the maximum likelihood estimator is defined as 
\begin{align}
\tilde{g}_{\text{\tiny MLE}}: \left.\frac{\partial \mathcal{L}}{\partial g}\right|_{\tilde{g}_{\text{\tiny MLE}}}=0;
\end{align}
one should also ensure that the second derivative is negative at this point to ensure a maximum and not a minimum. For a displaced Gaussian beam $p'(x|g)=p(x-\lambda g)$, it is easy to show that 
\begin{align}
\tilde{g}_{\text{\tiny MLE}}=\frac{1}{\lambda N}\sum_i x_i-\frac{x_0}{\lambda} .
\label{ideal_MLE}
\end{align}
The variance of this statistic can simply be propagated from the variance of each $x_i$ (note $x_0$ is a constant and has zero variance), and is given by $\text{Var}(\tilde{g}_{\text{\tiny MLE}})=\sigma^2/(N\lambda^2)$. By the efficiency of MLE we can then infer the Fisher information. Alternatively, what we call the `prior' Fisher information can be directly calculated 
\begin{align}
F_x:&=\int \frac{(\partial_gp'(x))^2 }{p'(x)}dx=\frac{\lambda^2}{\sigma^2}.
\label{ideal_F}
\end{align}
The subscript $x$ denotes an assumption of infinite detector resolution. The Fisher information is the canonical measure of metrological performance. A higher Fisher information indicates the possibility of a lower variance in the estimate of an unknown quantity. Efficient estimation strategies such as MLE can reach the ultimate limit in precision~\cite{Fisher1925}.

We imagine $\lambda,\sigma$ and $x_0$ to be parameters that may or may not be under the control of the experimenter. In the case of an ideal detector (with infinite resolution), clearly the ratio of $\lambda$ to $\sigma$ should be made as high as possible. Note that alignment is only relevant for data processing (the estimator~\eqref{ideal_MLE}): the expected performance (the Fisher information~\eqref{ideal_F}) does not depend on $x_0$. 

An alternative metric often used to characterise the performance of beam displacement experiments~\cite{BarnettFabreMaitre2003} or precision measurements more generally, is the signal-to-noise ratio (SNR). We call the `prior' SNR
\begin{equation}
R_x:=\frac{|\langle x \rangle-x_0|}{\sqrt{\langle x^2\rangle-\langle x \rangle^2}}=\frac{\lambda g}{\sigma}.\end{equation}

This quantity is defined as the magnitude of the ratio of the average displacement to the standard deviation (i.e. a function of the moments of $x$)~\footnote{Some authors define the signal-to-noise ratio as $\sqrt{N}R_x$.}. In fact for the Gaussian model \eqref{gaussian_model} considered here 
\begin{align}
R_x = g \sqrt{F_x}.
\end{align}
As we show in Sec.~\ref{split_snr}, $R_x$ is also the \emph{first order approximant} (up to a factor of $\sqrt{2/\pi}$) to $R_n$, the (posterior) signal to noise ratio of a split detector~\cite{BarnettFabreMaitre2003}. 

For ideal detection, when comparing the performance under two different values of $\lambda$, the `gain' for any fixed value of $g$ is the same whether one uses the square root of the prior Fisher information
\begin{align}
\sqrt{\frac{F_x[p(x-\lambda_1 g)]}{F_x[p(x-\lambda_2 g)]}}=\frac{\lambda_1}{\lambda_2}
\end{align}
or if one uses the prior SNR
\begin{align}
\frac{R_x[\lambda_1]}{R_x[\lambda_2]}=\frac{\lambda_1}{\lambda_2}.
\end{align}
A similar result is found for two values of the standard deviation $\sigma_1$ and $\sigma_2$. It can thus be tempting to use $R_x$, as it is far easier to work with. Caution must be employed, however, to ensure that nothing is being lost by reverting to the simpler figure of merit. As we shall see, when one uses a split detector, the two figures can behave very differently.
\section{High resolution detection}
\label{high_res}
The maximum likelihood analysis above will fail to apply in a real situation, where some sort of pixelation is expected. The Fisher information under pixelation is surprisingly robust against reduction in the spatial resolution of the detector~\cite{KneeGauger2014}. If we assume that the detector rounds the result of an ideal measurement to the nearest integer multiple of $r$ (the pixel width), this can be modelled by an appropriate transformation of the probability density function into a probability mass function
\begin{align}
p(x)\rightarrow \text{Pr}(n)=\int_{(n-1/2)r}^{(n+1/2)r} p(x)dx.
\end{align}
Here $n$ is an integer that indexes the pixels.

The likelihood is maximised with
\begin{equation}
\tilde{g}_{\text{\tiny MLE}}= \frac{r\sum_i n_i}{\lambda N}-\frac{x_0}{\lambda}.
\label{approx_unproblematic}
\end{equation}
This has a similar form to the unpixelated case~\eqref{ideal_MLE}. 

The shift $g\lambda+x_0$ can be split into an integer multiple of $r$ plus a remainder which becomes negligible as $r\rightarrow0$. In this regime of high resolution, the shift in $x$ is a `shift in $n$' and then $\text{Var}(n_i)\approx \frac{\sigma^2}{r^2}+\frac{1}{12}$~\cite{variance_discrete_normal}, and the variance of the maximum likelihood estimator is $ \text{Var}(\tilde{g}_{\text{\tiny MLE}} )\approx\frac{1}{\lambda^2N}(\sigma^2+\frac{r^2}{12})$ which is independent of $g$. 
The Fisher information penalty due to pixelation is the same for any $g$ and any $x_0$ as long as $r$ is small.~\cite{KneeGauger2014}. 
\section{Split detection}
\label{split}
In the opposite limit to high resolution detection,  we have split detection: when $r$ becomes large enough, only two pixels will be relevant. %
We imagine the split detector to consist of two pixels - one for positive and another for negative values of $x$. Such a detector reduces each value to its sign. We assume the pixels have infinite extent and carry and index $n\in\{-1.+1\}$. The appropriate transformation on the probability density function is therefore
\begin{align}
p'(x)\rightarrow P(n)=\frac{1}{2}\left(1+n\,\text{erf}\left(\frac{g\lambda+x_0}{\sqrt{2}\sigma}\right)\right).
\end{align}
One must use the discrete probability mass function definition of the Fisher information to reach what we call the `posterior' Fisher information
\begin{align}
F_n &:= \sum_n \frac{(\partial_g P(n))^2}{P(n)}\\
&=\frac{2 \lambda ^2 e^{-\frac{(x_0+g \lambda) ^2}{\sigma ^2}}}{\pi  \sigma ^2 \left(1-\text{erf}\left(\frac{x_0+g \lambda }{\sqrt{2} \sigma }\right)^2\right)}.
\end{align}
Now if $\lambda$ and $\sigma$ are fixed, one should choose the alignment such that $x_0=-g\lambda$ for best performance. Then one recovers $F_n=2F_x/\pi$~\cite{PotzelbergerFelsenstein1993}. Shifting the detector relative to the centroid of $p(x)$ in this way does not affect the prior information $F_x$, but it can mitigate the information penalty up to a factor of $2/\pi$. The problem is that such realignment depends on the unknown quantity $g$, and is very difficult to achieve. An adaptive technique should be possible, on the other hand, and one can use the goal of a symmetric distribution in the split detector to guide the alignment. 

\begin{figure*}
\includegraphics[width=12cm]{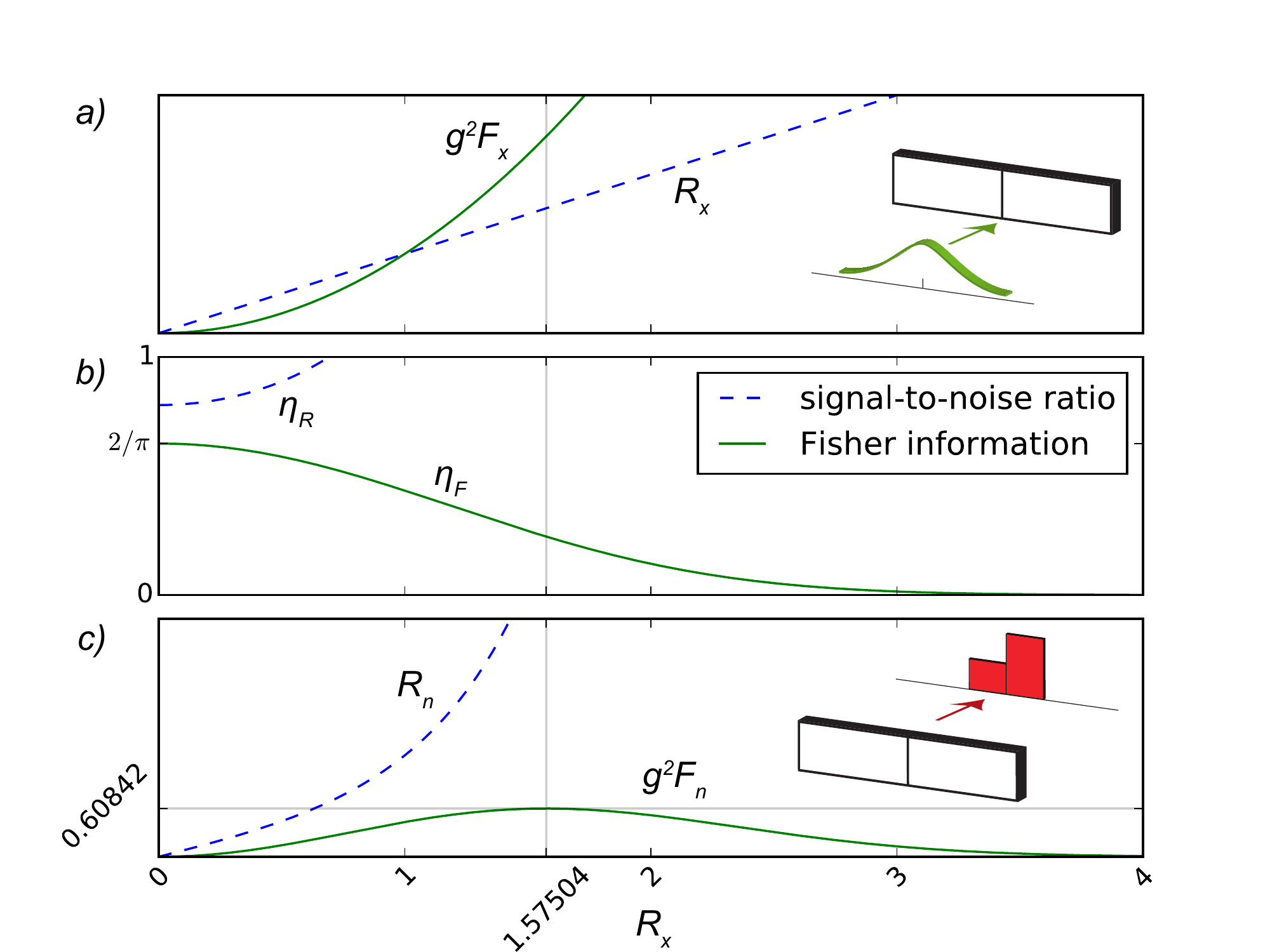}
\caption{\label{snrs_fi}(Color online) a) Prior quantities: the signal-to-noise-ratio $R_x$ (dashed, blue) and the scaled Fisher information $g^2F_x$ (solid, green) as a function of prior signal-to-noise $R_x$, assuming an ideal detector (note their monotonic relation to one another); b) Information penalty or `transmittance' $\eta_F$ of Fisher information (solid, green) or $\eta_R$ of signal-to-noise (dashed, blue) at the split detector when $x_0=0$; c) Posterior quantities: the signal-to-noise ratio $R_n$ (dashed, blue) and scaled Fisher information $g^2F_n$ (solid, green) relating to a well aligned ($x_0=0$) split detector. The measures are no longer monotonically related. All quantities are dimensionless. }
\end{figure*}

Another approach is to fix $x_0=0$; i.e. perfectly balance the detector before the unknown shift is introduced~\cite{FabreFouetMaitre2000}. Then one has
\begin{align}
g^2F_n=\frac{2}{\pi}\frac{ R_x^2 e^{-R_x^2}}{1-\text{erf}\left(\frac{R_x}{\sqrt{2}}\right)^2}.
\label{ggFn}
\end{align}
The dependence of $g^2F_n$ and $g^2F_x$ on $R_x$ is shown in Figure~\ref{snrs_fi}.
There are two competing behaviours as $R_x$ is increased: the prior information increases, but the information penalty of the split detector becomes more severe (the centroid of the Gaussian becomes further from the pixel boundary). 
Let us consider the Fisher information penalty or `transmittance' due to the split detector when $x_0=0$:
\begin{align}
\eta_F&:=\frac{F_n}{F_x}=\frac{2}{\pi}\frac{e^{-R_x^2}}{1-\text{erf}\left(\frac{R_x}{\sqrt{2}}\right)^2}.
\end{align}
Notice that this function is decreasing in $R_x$. It attains the upper bound of $2/\pi$ in the limit $R_x\rightarrow 0$.  Figure \ref{snrs_fi} shows that as the prior SNR increases, to first order the transmission of information is constant. Around $R_x=1$ the transmittance $\eta_F$ decreases linearly in $R_x$. These results are in fact entirely intuitive. The information about the shift must tend toward zero as $R_x\rightarrow\infty$, because in the limit of a large prior SNR, every click lands in the right pixel (say), and any measure on the set of possible values of $g$ becomes infinite, and it becomes impossible to make a nontrivial estimate. The Fisher information transmittance $\eta_F$ captures this mathematically. By contrast, attempting to define a signal-to-noise transmittance $\eta_R :=R_n/R_x$ leads to a quantity that exceeds unity, as shown in Figure~\ref{snrs_fi}.

By inspection of ~\eqref{ggFn}, when one has control over $R_x$, (i.e. over $\lambda$ or $\sigma$ or both) there will be an optimum choice to maximise $g^2F_n$ given split detection. The optimum value is $R_x^* =1.57504\ldots$~\footnote{The $R_x$ that maximises $g^2F_n$ is found by solving $\frac{\partial}{\partial R_x}\left(\frac{ R_x^2 e^{-R_x^2}}{1-\text{erf}\left(\frac{R_x}{\sqrt{2}}\right)^2}\right)=0$ and ensuring the second derivative is negative at this point}. This optimum point, well outside of the linear regime of the detector, sets a maximum achievable Fisher information of $F_n^*\approx 0.60842/g^2$. This is in contrast with high resolution detection where the information a) is independent of $g$ and b) can be freely increased by scaling $\lambda$ or $\sigma$. Note that adaptive realignment of a split detector effectively reproduces the high resolution situation, up to the factor of $2/\pi$. 
\section{Saturating the Cram\'{e}r-Rao bound}
\label{crb}
\begin{figure*}
\includegraphics[width=12cm]{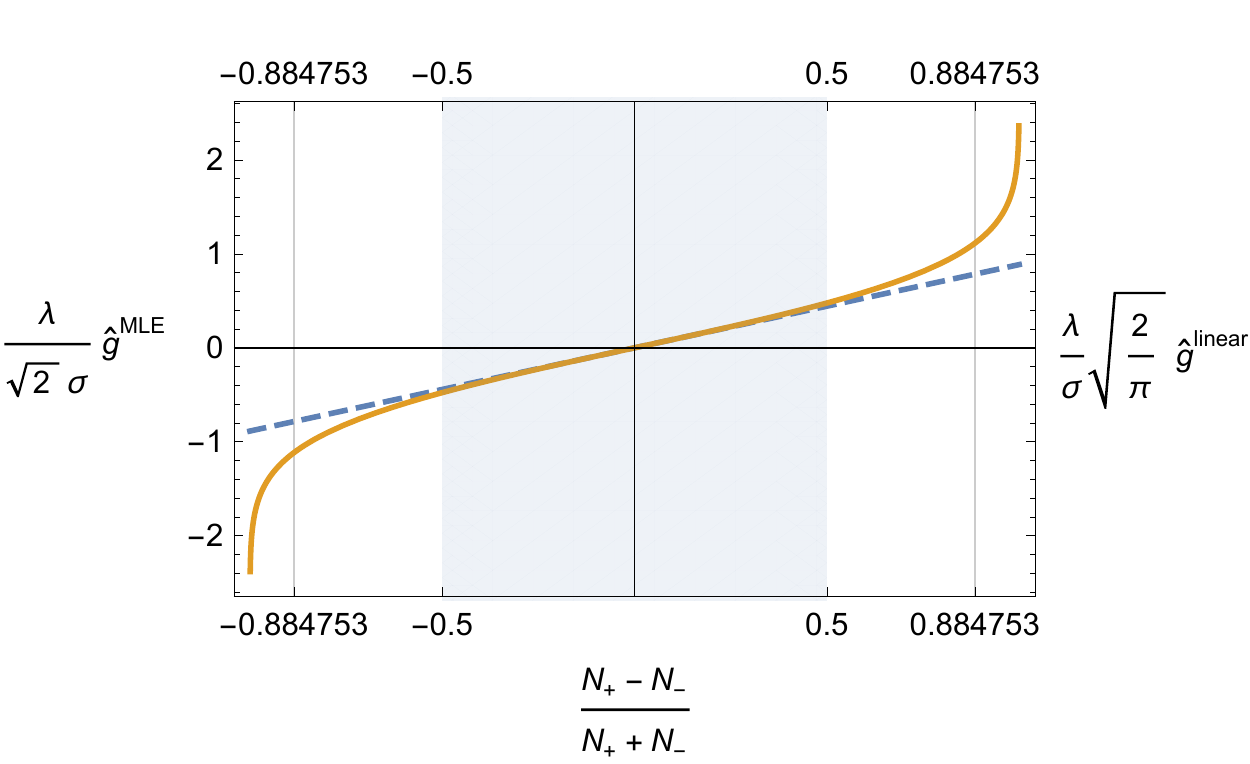}
\caption{(Color online) Dependence of two estimators on the observed normalised difference in counts at a well aligned ($x_0=0$) split detector. The blue, dashed line is proportional to the simple linear estimator, which can become significantly biased when the difference in counts is outside of the linear regime (shaded area). It is a first order approximant to the solid, orange curve, which is proportional to the maximum likelihood estimator. The latter is always unbiased and has unbeatable performance (mean squared error) when the number of trials is large. The optimum normalised difference in counts 0.884753\ldots balances the prior Fisher information $F_x$ with the information penalty $\eta_F$ to achieve the optimum posterior Fisher information $F_n$. This can be achieved by modulating $\lambda$ or $\sigma$.}
\label{inverf}
\end{figure*}

Maximum likelihood estimation is an efficient technique for saturating the  Cram\'{e}r-Rao bound. Often the estimation procedure is complicated enough the necessitate a numerical approach. For certain scenarios, however, an analytic closed form expression for the estimator can be calculated. This makes the computation of the estimate very easy, and can also provides insights into experiment design. 
For split detection, the likelihood function to be maximized is 
\begin{align}
\mathcal{L}(g|n_i)=\left[\frac{N_+!+N_-!}{N_+!N_-!}\right]P(+1)^{N_+}P(-1)^{N_-}
\end{align}
where $N_\pm$ is the number of clicks in each pixel. Now
\begin{align}
0&=\left.(\partial_g\log \mathcal{L})\right|_{\tilde{g}_{\text{\tiny MLE}}}\\
&=\partial_g \left( N_+\log\left[\frac{1}{2}+\frac{1}{2}\text{erf}\left(\frac{x_0+g\lambda}{\sqrt{2}\sigma}\right)\right]\right.\\
 &+\left. N_-\log\left[\frac{1}{2}-\frac{1}{2}\text{erf}\left(\frac{x_0+g\lambda}{\sqrt{2}\sigma}\right)\right] \right)_{\tilde{g}_{\text{\tiny MLE}}}\\
&=\left(\frac{N_+}{\frac{1}{2}+\frac{1}{2}\text{erf}\left(\frac{x_0+\tilde{g}\lambda}{\sqrt{2}\sigma}\right)}-\frac{N_-}{\frac{1}{2}-\frac{1}{2}\text{erf}\left(\frac{x_0+\tilde{g}\lambda}{\sqrt{2}\sigma}\right)}\right).
\end{align}
We divided by $e^{-\frac{(x_0+\tilde{g}\lambda)^2}{2\sigma^2}}$, a factor which tends to zero when $\tilde{g}\rightarrow \pm \infty$, extrema corresponding to the minimum likelihood. We also discarded the binomial factor, which became an additive constant under the action of the logarithm. Preceding with 
\begin{align}
\frac{N_-}{\frac{1}{2}-\frac{1}{2}\text{erf}\left(\frac{x_0+\tilde{g}\lambda}{\sqrt{2}\sigma}\right)}=\frac{N_+}{\frac{1}{2}+\frac{1}{2}\text{erf}\left(\frac{x_0+\tilde{g}\lambda}{\sqrt{2}\sigma}\right)}
\end{align} 
one arrives at the maximum likelihood estimator
\begin{align}
\tilde{g}_{\text{\tiny MLE}}=\frac{\sqrt{2}\sigma}{\lambda}\text{inverf}\left(\frac{N_+ -N_-}{N_+ +N_-}\right)-\frac{x_0}{\lambda}
\end{align}
Note the difference to the high resolution estimator~\eqref{approx_unproblematic}-- here we must take account of $\sigma$. The maximum likelihood estimator is approximated, when the argument of the inverf is small, by
\begin{align}
\tilde{g}_{\text{\tiny MLE}}\approx\tilde{g}_{\text{linear}}=\frac{\sqrt{\pi}\sigma}{\sqrt{2}\lambda}\left(\frac{N_+ -N_-}{N_+ +N_-}\right)-\frac{x_0}{\lambda}.
\end{align}
In fact this approximation is good until the argument approaches around one half. The departure of the maximum-likelihood estimate from a simple linear estimate is shown in Figure~\ref{inverf}.  
While the simple linear estimator becomes biased, the MLE continues to provide an unbiased estimate in the nonlinear regime, although of course performance is impacted.  Interestingly, when $x_0=0$, the optimum value for $(N_+-N_-)/(N_++N_-)$ is 0.884753\ldots which follows from the optimum $R_x$. 
\section{Split detection SNR}
\label{split_snr}
As is manifest from Figure~\ref{snrs_fi}, $R_x$ breaks down as a useful measure of precision for split detectors when it becomes too large. In this section we investigate whether a faithful application of the concept of signal-to-noise ratio to the split detector:
\begin{align}
R_n:=\frac{|\langle n \rangle|}{\sqrt{\langle n^2\rangle - \langle n \rangle^2}}&=\frac{\text{erf}\left(\frac{x_0+g\lambda}{\sqrt{2}\sigma}\right)}{1-\text{erf}\left(\frac{x_0+g\lambda}{\sqrt{2}\sigma}\right)^2}
\end{align}
can provide a more useful measure. When $x_0=0$ we have
\begin{align}
R_n&=\frac{\text{erf}\left(\frac{R_x}{\sqrt{2}}\right)}{1-\text{erf}\left(\frac{R_x}{\sqrt{2}}\right)^2}.
\end{align}
When $R_x\ll\sqrt{2}$, we have
\begin{align}
R_n \approx \sqrt{\frac{2}{\pi}}R_x
\end{align}
but otherwise $R_n$ is not monotonically related to $F_n$. The differences between $R_x$ and $F_n$ are not therefore attributable to to the failure of linearity in $R_x$, because the difference in trend between $R_n$ and $F_n$ is even more pronounced (see Figure~\ref{snrs_fi}) away from the linear regime. Since $R_n\rightarrow\infty$ as $R_x\rightarrow\infty$, it is therefore not advisable to use $R_n$ as a measure of precision. Even when there is a vanishing amount of information about $g$ available, $R_n$ reports an arbitrarily high signal-to-noise ratio.

This serves as a warning against using $R_n$ or $R_x$ as figures of merit, unless $R_x\ll\sqrt{2}$. When that condition is satisfied, the SNR is a good figure of merit and qualitatively captures the performance as measured by the Fisher information. Otherwise the Fisher information should be preferred because it takes the nonlinear relationship between what is measured ($n$) and what is estimated ($g$) into account.

\section{Application: weak-value amplification}
\label{wva}
Weak-value amplification is a probabilistic signal amplification method which operates via quantum interference. Its discovery arose from a time symmetric approach to quantum theory~\cite{AharonovAlbertVaidman1988}. By weakly coupling two degrees of freedom, usually named `system' and `meter', and then postselecting on a unlikely outcome of a subsequent strong measurement of the system, an unexpected change is seen in the meter~\cite{KneeCombesFerrie2014}. The meter evolves as if it interacted with a `system' of much higher energy. 
 
The split detector is the detector of choice for WVA experiments in optical systems:~\cite{HostenKwiat2008,DixonStarlingJordan2009}, where the transverse state of a light beam acts as the `meter' and records (for example) the polarisation or which-path state of the beam. By employing a weak-value protocol, the beam is shifted much further than usual, albeit probabilistically. 

Moreover, the signal-to-noise ratio $R_x$ is usually preferred in these experiments for its being an `intuitive concept'~\cite{Kedem2012}, leading to practical estimation strategies~\cite{JordanMartinez-RinconHowell2014}. In fact this situation has been considered by Str\"ubi and Bruder~\cite{StrubiBruder2013} who calculated an approximation to $F_n$ in the linear regime. By contrast we imagine that MLE could be employed to unlock the non-linear regime of the split detector, and to guarantee unbiased estimates.  As we have shown, the Fisher information $F_n$ is then not simply related to $R_x$.  

To allow for a discussion of WVA we imagine an ancillary `system' prepared in an eigenstate of quantum mechanical observable $\mathbf{A}$ with eigenvalue $\lambda$, coupled impulsively to the transverse momentum of our beam with Hamiltonian $H = g\mathbf{A}\hat{k}_x$. Since  $\hat{k_x}$ describes the transverse momentum of the beam, and is the generator of translations in $x$, this leads to $p(x-g\lambda)$ as above. Here we assume the initial state is such that the largest magnitude eigenvalue $\lambda_*$ is selected. When the coupling is weak $g\rightarrow 0$, however, and the ancillary system is prepared in an arbitrary state $|i\rangle$, impulsively coupled and then found to be final state $|f\rangle$,  one substitutes the largest eigenvalue $\lambda_*$ with 
\begin{align}
A_w :=\frac{\langle f |\mathbf{A} |i\rangle}{\langle f| i \rangle}
\end{align} 
the `weak value'~\cite{AharonovAlbertVaidman1988} (here taken to be a real number). This quantity can become much larger in magnitude than $\lambda_*$ by tuning $|i\rangle$ and $|f\rangle$. Crucially, however, there is \emph{no associated increase in the net Fisher information} $NF_x$. This is because the technique only succeeds with a typically small probability $q=|\langle f|i\rangle|^2$, leading to only $qN$ successful runs. The reader is referred to Ref.~\cite{DresselMalikMiatto2014} for further details.

The square of the `gain' serves as a measure of the relative performance of weak-value amplification. We will once more set $x_0=0$. Now
\begin{align}
\frac{qNF_n^{\text{wva}}}{NF_n^{\text{std}}} =& \frac{qF_{x}^{\text{wva}}}{F_{x}^{\text{std}}}\frac{\eta_F(gA_w/\sigma)}{\eta_F(g\lambda_*/\sigma)} \\
=&\frac{|\langle f |\mathbf{A} |i\rangle|^2}{\lambda_*^2}\frac{\eta_F(gA_w/\sigma)}{\eta_F(g\lambda_*/\sigma)}.
\end{align}
The first factor can be taken towards unity, but never exceeds it~\cite{KneeGauger2014}. The second factor is the ratio of transmittances, or Fisher information penalties, due to the split detector. It is to first order unity, but in general decreasing in $A_w/\lambda_*$. Either $A_w=\lambda^*$ or $A_w\gg \lambda^*$ is necessary for the first factor to approach unity~\cite{KneeGauger2014}. In the first case the weak-value technique reduces to largest-eigenvalue technique and no amplification is seen. In the second case there is a large amplification, but this is sufficient to ensure the second factor is \emph{less} than unity. So real WVA is strictly worse than a maximum eigenvalue method with a split detector -- this conclusion is consistent with Str\"ubi and Bruder's claim that one should operate away from the regime of large amplification~\cite{StrubiBruder2013}.

\section{discussion}
\label{discussion}
We have made a formal comparison of Fisher information and signal-to-noise before and after a split detector. Considering maximum-likelihood estimation as an unbiased and efficient estimation strategy, we found that the signal-to-noise ratio becomes misleading as a measure of precision outside of the linear response regime of the detector.  Outside of this regime, simple linear estimation becomes biased. Both of these problems can be overcome by using MLE, which is unbiased across the entire regime of the split detector and efficient (meaning its mean squared error is given by $F_n$).

Furthermore we discussed ways to optimize the use of split detectors. Firstly, adaptive realignment of the detector (taking $x_0\rightarrow -\lambda g$) is a challenging but rewarding technique which can in principle recover the ideal Fisher information $F_x$ up to a multiplicative constant of $2/\pi$. A more realistic technique is to balance the detector when there is no signal $x_0=0$: then one can achieve higher performance (i.e. an unbiased estimate with a lower mean squared error) with the same number of photons by modulating $\lambda/\sigma$ such that the detecor is operating at an optimum point $R_x^*$ outside the linear regime. In optics this can be achieved by changing the spot size of the beam. In so doing, instead of operating close to the limit of the linear response region $N_+-N_-=0.5(N_++N_-)$ and $R_x\approx0.675$,  one can make $R_x\approx 1.57504$ and so achieve approximately a 248\% improvement in Fisher information. The advantage over operating deep within the linear regime (instead of at the limit) will be even higher. 

We also showed that in the absence of any other technical problems, the use of split detection implies that (real) weak-value amplification has a strictly worse performance (lower posterior Fisher information) than a standard technique. This is similar to the problem of nonlinear bias in the weak-value linear estimator~\cite{KneeCombesFerrie2014}, in that it can be reduced by controlling $\sigma\rightarrow\infty$ (equivalently when $g\rightarrow0$) so that the two strategies performance approaches parity. 

\section{acknowledgements}
We thank Eliot Bolduc for a careful reading of this manuscript.
\appendix

\bibliography{/Users/georgeknee/Documents/NTT_BRL_2014_2015/paper_library/gck_full_bibliography}

\begin{thebibliography}{10}

\bibitem{GerlachStern1922}
W.~{Gerlach} and O.~{Stern},
\newblock Zeitschrift fur Physik {\bf 9}, 353 (1922).

\bibitem{Young1804}
T.~Young,
\newblock Philosophical Transactions of the Royal Society of London {\bf 94}, 1
  (1804).

\bibitem{Davisson1928}
C.~Davisson,
\newblock Bell System Technical Journal, The {\bf 7}, 90 (1928).

\bibitem{HostenKwiat2008}
O.~Hosten and P.~Kwiat,
\newblock Science {\bf 319}, 787 (2008).

\bibitem{AielloWoerdman2008}
A.~Aiello and J.~P. Woerdman,
\newblock Opt. Lett. {\bf 33}, 1437 (2008).

\bibitem{SunEdgarBowman2013}
B.~Sun {\em et~al.},
\newblock Science {\bf 340}, 844 (2013).

\bibitem{Wolfowitz1965}
J.~Wolfowitz,
\newblock Theory of Probability \& Its Applications {\bf 10}, 247 (1965),
  http://dx.doi.org/10.1137/1110029.

\bibitem{Wald1949}
A.~Wald,
\newblock The Annals of Mathematical Statistics , 595 (1949).

\bibitem{Van-Trees1968}
H.~Van~Trees,
\newblock {\em Detection, Estimation, and Modulation Theory}, Detection,
  Estimation, and Modulation Theory No.  pt. 1 (Wiley, 1968).

\bibitem{Fisher1925}
R.~A. Fisher,
\newblock Mathematical Proceedings of the Cambridge Philosophical Society {\bf
  22}, 700 (1925).

\bibitem{BarnettFabreMaitre2003}
S.~Barnett, C.~Fabre, and A.~Maitre,
\newblock The European Physical Journal D - Atomic, Molecular, Optical and
  Plasma Physics {\bf 22}, 513 (2003).

\bibitem{Note1}
Some authors define the signal-to-noise ratio as $\protect \sqrt {N}R_x$.

\bibitem{KneeGauger2014}
G.~C. Knee and E.~M. Gauger,
\newblock Phys. Rev. X {\bf 4}, 011032 (2014).

\bibitem{variance_discrete_normal}
R.~Israel,
\newblock Estimating the variance of a discrete normal distribution,
\newblock MathOverflow, 2015,
  http://mathoverflow.net/questions/178964/estimating-the-variance-of-a-discrete-normal-distribution.

\bibitem{PotzelbergerFelsenstein1993}
K.~P{\"o}tzelberger and K.~Felsenstein,
\newblock Journal of Statistical Computation and Simulation {\bf 46}, 125
  (1993).

\bibitem{FabreFouetMaitre2000}
C.~Fabre, J.~B. Fouet, and A.~Ma\^{i}tre,
\newblock Opt. Lett. {\bf 25}, 76 (2000).

\bibitem{Note2}
The $R_x$ that maximises $g^2F_n$ is found by solving $\protect \frac {\partial
  }{\partial R_x}\left (\protect \frac { R_x^2 e^{-R_x^2}}{1-\protect \text
  {erf}\left (\protect \frac {R_x}{\protect \sqrt {2}}\right )^2}\right )=0$
  and ensuring the second derivative is negative at this point.

\bibitem{AharonovAlbertVaidman1988}
Y.~Aharonov, D.~Z. Albert, and L.~Vaidman,
\newblock Phys. Rev. Lett. {\bf 60}, 1351 (1988).

\bibitem{KneeCombesFerrie2014}
G.~C. Knee, J.~Combes, C.~Ferrie, and E.~M. Gauger,
\newblock (2014), 1410.6252v1.

\bibitem{DixonStarlingJordan2009}
P.~Dixon, D.~J. Starling, A.~N. Jordan, and J.~C. Howell,
\newblock Phys. Rev. Lett. {\bf 102}, 173601 (2009).

\bibitem{Kedem2012}
Y.~Kedem,
\newblock Phys. Rev. A {\bf 85}, 060102 (2012).

\bibitem{JordanMartinez-RinconHowell2014}
A.~N. Jordan, J.~Martinez-Rincon, and J.~C. Howell,
\newblock Phys. Rev. X {\bf 4}, 011031 (2014).

\bibitem{StrubiBruder2013}
G.~Str\"ubi and C.~Bruder,
\newblock Phys. Rev. Lett. {\bf 110}, 083605 (2013).

\bibitem{DresselMalikMiatto2014}
J.~Dressel, M.~Malik, F.~M. Miatto, A.~N. Jordan, and R.~W. Boyd,
\newblock Rev. Mod. Phys. {\bf 86}, 307 (2014).

\end{thebibliography}
\bibliographystyle{h-physrev}

\end{document}